**Autonomous Generation of Metamaterial Databases Based on Multimodal Agents**

*Shilong Qin[1,†], Zhicai Yu[1,†], Xuan Zheng[1,†], Yan Zhang[1], Aodi Yang[1],Kezhan Zhao[1], Qi Cheng Chen[1], Jian Wei You [1,*] and Tie Jun Cui[1,2,*]*

[1] State Key Laboratory of Millimeter Wave, Southeast University, Nanjing 210096, China

[2] Center of Metamaterials, Suzhou National Laboratory, Suzhou, 215000, China.

[†] These authors contributed equally to this work.

*Corresponding authors: jvyou@seu.edu.cn, tjcui@seu.edu.cn

## Abstract

Artificial intelligence (AI) is revolutionizing material research and discovery. However, its development in metamaterials is bottlenecked by a shortage of high-quality and executable structure-response databases, which are locked within scientific literatures as a mixture of text and images. Converting the rapidly growing body of scientific literatures into executable and reusable databases for machine-driven discovery is still a fundamental challenge. Here, we propose MetaDataGenAgent, a multimodal multi-agent framework that autonomously converts unstructured scientific literatures directly into metamaterial structure-response databases. MetaDataGenAgent establishes a complete literature-to-simulation pipeline through the coordinated operation of specialized agents for multimodal parameter extraction, physics-guided validation, topology-aware structural analysis, and solver-executable encoding. The framework introduces a closed-loop plan-execute-reflect mechanism that enables dynamic task decomposition, iterative validation, and feedback-driven model construction. Experimental results validate that MetaDataGenAgent can generate high-fidelity structure-response data for representative meta-atoms, which are further used to realize diverse electromagnetic functions, including far-field beam deflection, near-field holographic imaging and topologically protected surface-wave transport. By establishing an autonomous route from scientific literatures to AI-ready databases, the framework provides a general and efficient strategy that could be extended to a broad range of data-scarce scientific domains, including photonics, materials science, chemistry, computational science, and scientific automation.

## Introduction

Artificial intelligence (AI) is reshaping materials research by accelerating prediction, design, optimization, and discovery. Recent advances in large language models (LLMs), multimodal foundation models, and vision-language models (VLMs) have shown growing potentials in scientific literature understanding, materials discovery, molecular and protein design, and code generation[1–8]. These capabilities are particularly relevant to electromagnetic (EM) metamaterials and metasurfaces, where structural geometry, material composition, and EM response are strongly coupled. Over the past decade, metamaterials and metasurfaces have enabled versatile control of EM waves and have supported a wide range of functions, including beam steering, focusing, holographic imaging, programmable scattering, intelligent wireless environments, and topological wave transport[9–22]. Despite these advances, the use of AI for metamaterial design remains fundamentally limited by the scarcity of high-quality, reproducible, and executable structure-response databases[23-24].

The database bottleneck originates from the way metamaterial knowledge is generated, reported, and stored. Most experimentally or numerically validated designs are described in scientific papers through a combination of text descriptions, structural schematics, material parameters, simulation settings, and measured or simulated responses. Although these reports contain rich design knowledge, they are not directly machine-readable, standardized, or simulation-ready. Hence, constructing a usable database entry from a published design typically requires researchers to identify geometric parameters, interpret structural schematics, assign materials, rebuild the model in a full-wave solver, define boundary and excitation conditions, and verify the resulting EM response. This process is labor-intensive, time-consuming, and highly dependent on expert experience, which limits both the scalability and reproducibility of database construction. Consequently, a large body of validated metamaterial knowledge remains difficult to convert into reusable data assets for AI-driven design.

Multimodal foundation models offer a promising opportunity to address this challenge because they can jointly interpret textual and visual information from scientific documents[25-34]. However, transforming a published metamaterial design into an executable simulation model requires more than generic multimodal recognition. The necessary information is often dispersed across the main text, figure captions, structural diagrams, tables, and supplementary materials, and must be integrated into a complete, physically consistent, and solver-compatible modeling description. For full-wave EM simulation, this description should include geometric dimensions, material properties, layer configurations, boundary conditions, excitation settings, frequency ranges, and solver-specific implementation details. Even when these elements can be individually extracted, they must be cross-checked, completed, and organized according to EM modeling requirements before reliable structure-response data can be generated. Therefore, executable database construction requires an integrated framework that combines multimodal extraction, physics-guided validation, topology-aware structure analysis, and solver-executable encoding.

Current automated reconstruction and generative design approaches are not fully suited to this objective. Point-cloud-based reverse modeling methods rely on measured or scanned three-dimensional data and are difficult to apply to abstract structural schematics in publications. Latent generative models, including variational autoencoders and vector-quantized variational autoencoders, can generate geometric representations, but they generally lack explicit physical constraints and are not directly coupled to EM solvers[35–38]. Recent multimodal three-dimensional generation and computer-aided design methods have improved visual alignment and shape generation, but the resulting models often lack the semantic decomposition required for assigning materials, defining boundaries, or configuring simulations[39–41]. These limitations indicate that database construction for metamaterials should go beyond geometry generation. It must recover physically meaningful parameters, resolve structural topology, validate modeling completeness, and produce executable simulation scripts for standardized EM response calculation.

Here, we report MetaDataGenAgent, an orchestrator-driven multimodal agentic framework for constructing executable EM metamaterial databases from scientific literature. Instead of relying on a single foundation model or a fixed manual workflow, MetaDataGenAgent coordinates specialized agents and deterministic tools through a plan-execute-reflect process. The framework extracts geometric, material, and simulation-related information from text and figures, validates the completeness and physical consistency of the extracted parameters, analyzes structural topology when required, and encodes the validated description into solver-executable scripts. In this way, unstructured multimodal knowledge in the literature is transformed into simulation-ready structures and EM structure-response data. We validate MetaDataGenAgent using representative metamaterial systems with distinct physical mechanisms and modeling requirements, including discrete coding meta-atoms, Pancharatnam-Berry phase meta-atoms, anisotropic polarization-sensitive meta-atoms, and topological photonic unit cells. The generated responses show good agreement with published or benchmarked results. We further assemble the generated meta-atoms into functional metasurfaces for beam deflection, holographic imaging, EM focusing, and topologically protected surface-wave transport, and the agreement between simulations and experiments confirms their reliability for device-level EM design. Overall, MetaDataGenAgent establishes an automated literature-to-simulation route for addressing the database bottleneck in AI-driven metamaterial research and highlights how multimodal scientific knowledge can be transformed into executable data, thereby bridging the gap between published literature and AI-ready materials databases.

# Results

## Framework of MetaDataGenAgent

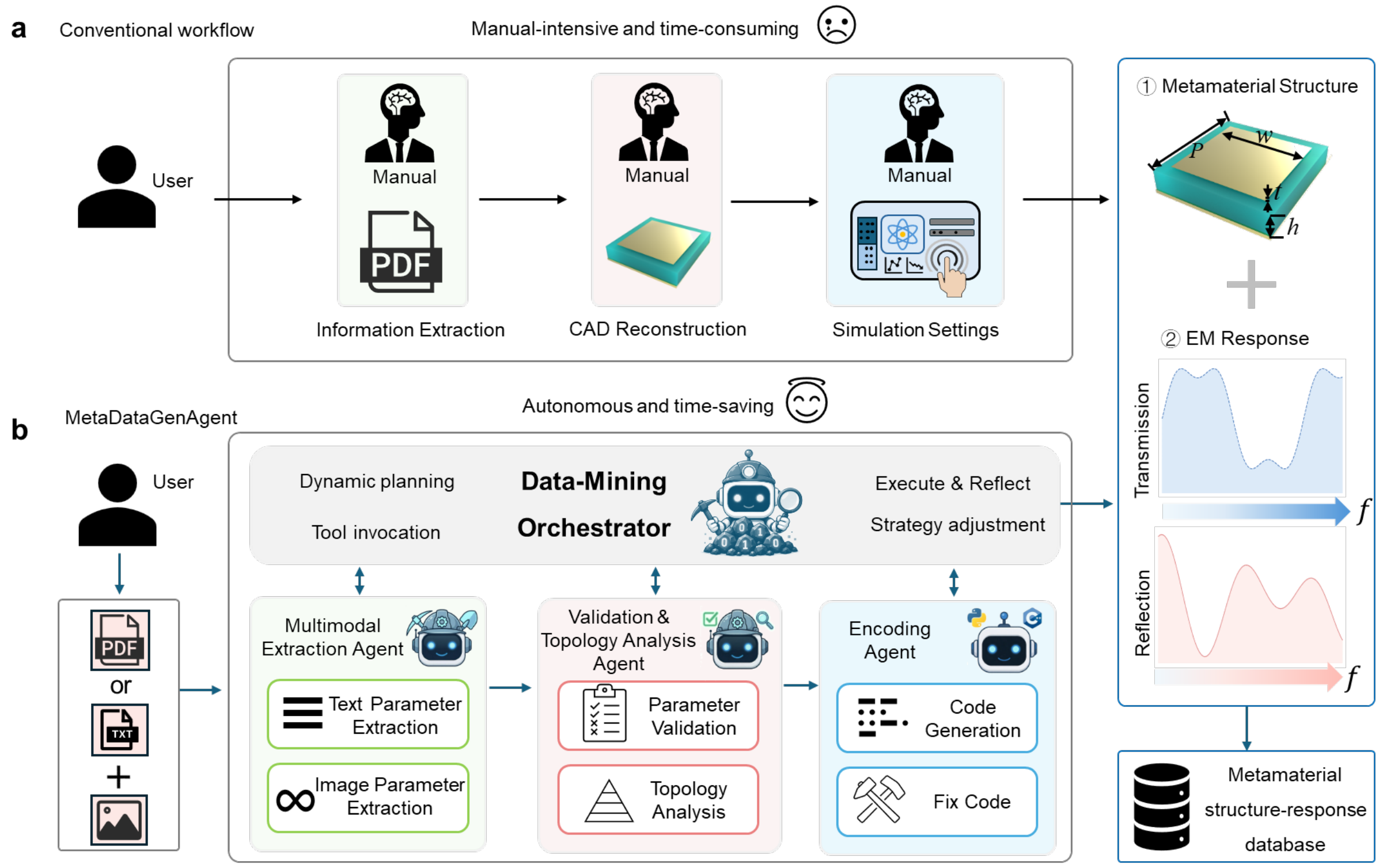

**Fig. 1| MetaDataGenAgent for metamaterial-database autonomous generation. a**, Conventional expert-driven workflow for constructing structure-response data from published metamaterial literatures. **b**, Proposed MetaDataGenAgent framework. An orchestrator manages the literature-to-database construction process through a plan-execute-reflect loop and coordinates three specialized agents for multimodal parameter extraction, physics-guided validation and topology analysis, and solver-executable encoding. This workflow transforms multimodal literature inputs into validated structure descriptions, executable simulation models, electromagnetic responses, and reusable database entries.

Executable structure-response databases for metamaterials are still commonly constructed through expert-driven manual reconstruction. As shown in **Fig. 1a**, a typical workflow begins with published papers, from which researchers identify structural geometry, material parameters, and simulation conditions. The reported design is then rebuilt in computer-aided design or full-wave simulation software, followed by material assignment, boundary-condition definition, excitation setup, frequency sweeping, and response verification. Although this procedure can produce reliable data, it is labor-intensive, difficult to standardize, and poorly scalable for mining the rapidly expanding metamaterial literature. To automate this process, we develop MetaDataGenAgent, an orchestrator-driven multimodal agentic framework that converts scientific literature into executable metamaterial database entries. As shown in **Fig. 1b**, MetaDataGenAgent accepts multimodal literature inputs, including full papers, structural figures, captions, and text descriptions, and transforms them into validated structure descriptions, solver-executable models, simulated electromagnetic responses, and reusable database entries. The workflow is coordinated by an orchestrator operating through a plan-execute-reflect loop. During execution, the orchestrator maintains the task state, dispatches specialized agents and deterministic tools, evaluates intermediate outputs, and triggers additional extraction, validation, topology analysis, or code repair when needed.

The framework is implemented through three specialized agents with complementary roles. The multimodal extraction agent retrieves geometric, material, and simulation-related information from textual and visual sources. The validation and topology analysis agent checks the extracted information for completeness, physical consistency, and structural executability, and further resolves topology for geometrically complex designs. The encoding agent translates the validated modeling description into solver-executable scripts and uses execution feedback to improve code reliability. By linking literature understanding, physics-guided validation, topology-aware modeling, and executable simulation, MetaDataGenAgent transforms published metamaterial designs into standardized structure-response data with reduced dependence on manual reconstruction.

**Agentic Coordination and Feedback Control in MetaDataGenAgent**

The internal agentic coordination mechanism of MetaDataGenAgent is illustrated in **Fig. 2**. The framework is organized around an orchestrator that manages the literature-to-database construction process through a plan-execute-reflect loop. For each input paper, the orchestrator initializes a task state from the extracted text, structural figures, captions, tables, and available supplementary information. It then decomposes the overall objective into executable subtasks and dispatches specialized agents and deterministic tools according to the current state. Intermediate outputs are evaluated and stored in runtime state memory, enabling the orchestrator to revise the task plan when information is incomplete, inconsistent, or not directly executable. This closed-loop mechanism allows MetaDataGenAgent to trigger additional extraction, validation, topology analysis, or code repair without restarting the entire construction process.

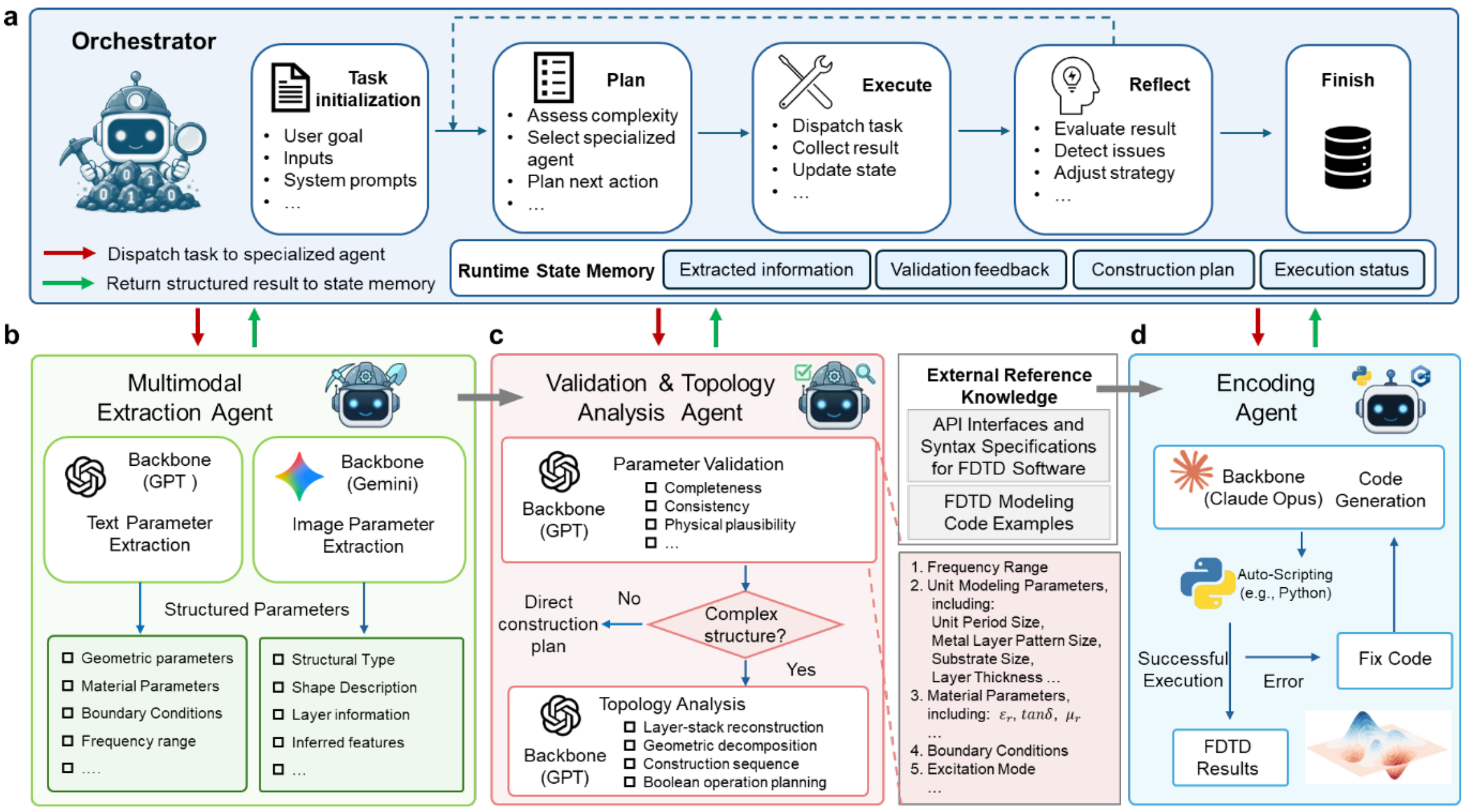

**Fig. 2| Agentic coordination and feedback control in MetaDataGenAgent. a**, Orchestrator manages the literature-to-database construction process through a plan-execute-reflect loop with runtime state memory. It initializes the task state from multimodal literature inputs, decomposes the overall objective into executable subtasks, dispatches specialized agents and deterministic tools, evaluates intermediate outputs, and triggers re-extraction, validation, topology analysis, and code repair. **b**, Multimodal extraction agent retrieves geometric, material, simulation-related, and structural information from textual and visual sources and fuses them into a structured parameter representation. **c**, Validation and topology analysis agent checks dimensional completeness, unit consistency, physical plausibility, and structural executability, and generates a topology-aware construction description for complex geometries. **d,** Encoding agent converts the validated modeling description into solver-executable scripts and uses execution feedback to improve model executability and code reliability.

Because the information required for EM modeling is often distributed across multiple parts of a paper, MetaDataGenAgent first constructs an initial design representation through coordinated textual and visual parsing. The textual branch extracts geometric dimensions, material properties, frequency ranges, boundary conditions, excitation settings, and other simulation-related descriptors from the main text, captions, tables, and supplementary files. In parallel, the visual branch analyzes structural schematics to recover topology, layer arrangement, spatial relationships, and geometric features that are difficult to define completely in text. The two information streams are then fused into a structured parameter set, which provides the basis for physics-guided validation and executable model construction.

Before script generation, the extracted representation is examined by the validation and topology analysis agent. This agent checks whether the parameter set satisfies the basic requirements for full-wave EM simulation, including dimensional completeness, unit consistency, physically reasonable material parameters, compatible boundary and excitation settings, and unambiguous correspondence between structural components and assigned materials. When missing or conflicting information is detected, the orchestrator uses the accumulated runtime state to guide re-extraction, correction, or parameter completion. For simple unit-cell geometries, the validated parameters can be directly organized into a construction plan. For more complex designs, the topology analysis module further decomposes the structure into layer stacks, geometric primitives, spatial relationships, and Boolean operations. This topology-aware intermediate representation reduces ambiguity in model reconstruction and improves the reliability of subsequent executable encoding.

The encoding agent converts the validated description into solver-executable scripts for full-wave simulation. Rather than relying on unconstrained code generation, the encoding process is grounded in solver application programming interface specifications, syntax requirements, and validated modeling examples. The generated script is executed through the corresponding solver interface to construct the model and calculate the EM response. If syntax errors, invalid parameters, missing objects, or failed simulation settings occur, the execution feedback is returned to the orchestrator and incorporated into the runtime state memory. The encoding agent then repairs the script according to the updated state and repeats execution when necessary. This feedback-guided encoding process improves the executability of the generated models and reduces the need for manual debugging.

After successful execution, the reconstructed geometry, material settings, simulation conditions, swept parameters, and calculated EM responses are organized into standardized structure-response entries. Through this agentic coordination process, MetaDataGenAgent integrates multimodal literature understanding, physics-guided validation, topology-aware modeling, and executable simulation into a unified literature-to-database construction framework. Detailed descriptions of the main orchestrator, specialized sub-agents, prompt design, tool interfaces, topology analysis strategy, and solver-specific encoding procedures are provided in Supplementary Notes 1-5.

**Role of multimodal extraction in executable response generation**

Reliable structure-response generation requires the extracted information to describe not only the numerical dimensions of a meta-atom, but also its structural topology and spatial arrangement. This requirement becomes particularly important when the electromagnetic response is governed by fine geometric features that are only partially described in text. To evaluate the role of visual information in executable database construction, we performed an ablation analysis using the cross-shaped anisotropic meta-atom shown in **Fig. 3**a. Two extraction pathways were compared under the same downstream validation and encoding procedure: a text-only pathway using information extracted from the source paper and a multimodal pathway combining textual extraction with visual structural analysis.

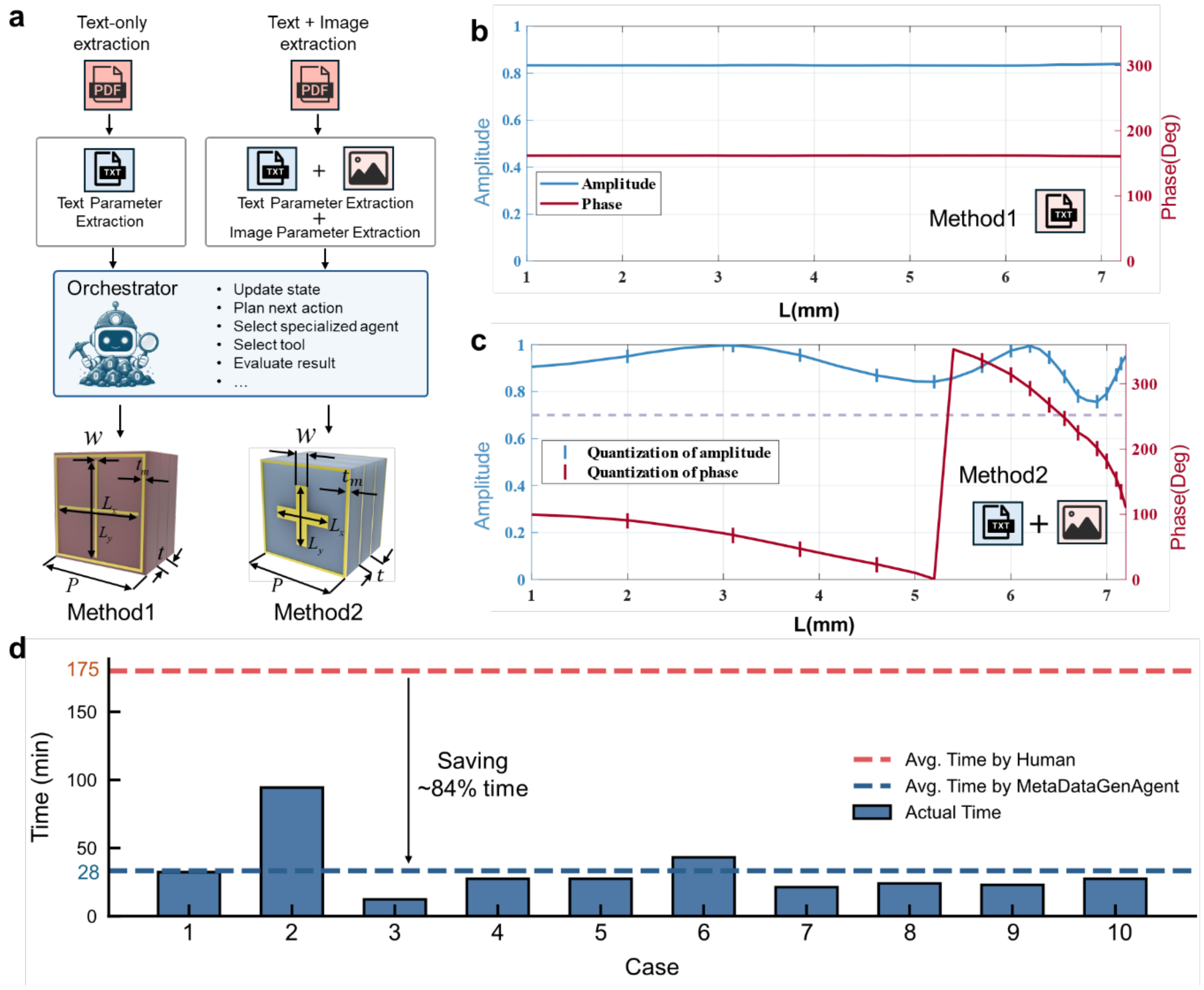

**Fig. 3| Ablation analysis of text-only and multimodal extraction for executable structure-response generation. a**, Comparison between the text-only and multimodal pathways for reconstructing a cross-shaped anisotropic meta-atom. The text-only pathway extracts parameters only from textual information, whereas the multimodal pathway combines textual parameter extraction with visual structural analysis. **b,c**, Full-wave parametric sweep results at 12.6 GHz for the meta-atom generated through the text-only and multimodal pathways, respectively. The arm length $L_x$is fixed. **d**, Modelling-efficiency benchmark on ten representative literature-derived cases. Cases 1-6 correspond to refs. 45-50, cases 7-9 correspond to refs. 42-44, and case 10 corresponds to ref. 14. The dashed red and blue lines denote the average human-expert modelling time and the average author-side modelling time, respectively, while the blue bars denote the case-specific modelling time of MetaDataGenAgent.

For this meta-atom, several global parameters can be obtained from the text, including the unit-cell period $P$, substrate thickness $t$, and metal thickness $t_m$. However, the linewidth $w$of the cross-shaped resonator is primarily specified by the structural schematic. As summarized in Table 1, the text-only pathway correctly extracts $P = 8$mm, $t = 6$mm, and $t_m = 0.2$mm, but assigns $w = 0.2$mm. This value corresponds to the metal thickness rather than the resonator linewidth. By contrast, the multimodal pathway recovers $w = 1$ mm, consistent with the original design. This comparison indicates that the dominant error in the text-only pathway arises from ambiguous structural interpretation before executable model generation. The impact of this parameter error is directly reflected in the simulated responses. Although the text-derived description can still be converted into a nominal three-dimensional model, the reconstructed geometry does not reproduce the expected tunable behavior. As shown in **Fig. 3b**, sweeping the arm length $L_y$ produces nearly unchanged transmission amplitude and phase responses. The incorrectly assigned linewidth therefore suppresses the response variation required for phase modulation and leads to an executable but physically unreliable structure–response entry.

When visual structural information is included, MetaDataGenAgent identifies the complete linewidth and topology of the cross-shaped resonator from the two-dimensional schematic and fuses this information with the material and simulation parameters extracted from text. The resulting multimodal description passes the subsequent physical-consistency checks and produces a more faithful executable model. As shown in **Fig. 3c**, sweeping $L_y$ from 1 to 7.2 mm leads to pronounced amplitude variation and a phase coverage close to $360°$. The discrete response points extracted from the simulated curves further indicate that the generated data can support phase quantization for metasurface design. These results demonstrate that visual information is not merely supplementary, but directly determines whether a literature-derived model can yield an executable and physically meaningful structure–response entry for geometrically complex meta-atoms.

We further quantified the modelling efficiency of MetaDataGenAgent using ten representative literature-derived cases, denoted as cases 1-10 in Fig. 3d. Cases 1-6 correspond to refs. 45-50[45-50], cases 7-9 correspond to refs. 42-44[42-44], and case 10 corresponds to ref. 14. The reported time includes only the construction of an executable electromagnetic model from the source publication, covering multimodal information extraction, parameter validation, geometry reconstruction and script generation, while excluding the subsequent electromagnetic simulation time. The red dashed line indicates the average modelling time required by human experts, the blue dashed line indicates the average modelling time in the authors' manual reproduction workflow, and the blue bars denote the case-specific modelling time required by MetaDataGenAgent. MetaDataGenAgent substantially reduces the modelling time relative to the human baseline across all cases, with even the longest run remaining nearly twofold shorter than the average human workflow. The increased time in this case mainly arises from missing or ambiguous parameters in the original publication, which required repeated reasoning, validation and parameter completion before an executable model could be produced. These results demonstrate that, by integrating multimodal fusion with physics-guided validation, MetaDataGenAgent improves the reliability and scalability of literature-derived electromagnetic model construction, thereby substantially accelerating the conversion of published designs into simulation-ready models for database generation. For the six literature-derived cases 1-6 used in the modelling-efficiency benchmark, the original simulation results reported in the source publications are compared with the corresponding MetaDataGenAgent reproductions in **Supplementary Fig. 2**. For the remaining cases 7-10, which cover additional representative meta-atoms and metasurface designs, the comparison between the literature-reported and MetaDataGenAgent-generated responses is provided in **Fig. 4**. Across all ten cases, the reproduced responses retain the main spectral features, phase variations and parameter-dependent trends of the original designs, further supporting the reliability and generalizability of the generated executable models.

**Table 1** | Comparison of structural parameters extracted by text-only and multimodal pathways with those reported in the original literature.

| Parameter | Reported value [[44]] (mm) | Text-only extraction (mm) | Multimodal extraction (mm) |
|---|---|---|---|
| $P$ | 8 | 8 | 8 |
| $t$ | 6 | 6 | 6 |
| $t_m$ | 0.2 | 0.2 | 0.2 |
| $w$ | 1 | 0.2 | 1 |

**High-Fidelity Reproduction and Generalization of MetaDataGenAgent**

To further evaluate whether the executable models generated by MetaDataGenAgent can reproduce the electromagnetic behaviours reported in the source literature, we examined four additional cases from the modelling-efficiency benchmark in **Fig. 3d**, denoted as cases 7-10. These cases complement the six literature-derived cases 1-6 shown in **Supplementary Fig. 2** and span distinct physical mechanisms, geometric topologies and simulation targets, including programmable phase coding, Pancharatnam-Berry (PB) phase modulation, anisotropic polarization-dependent responses and topological photonic band structures. Starting from the source publications or paired textual and visual inputs, MetaDataGenAgent extracted the relevant multimodal parameters, validated the modelling description, constructed executable electromagnetic models and generated the corresponding structure-response results, as summarized in **Fig. 4**. Detailed dimensional parameters of the generated meta-atoms are provided in **Supplementary Table 1**.

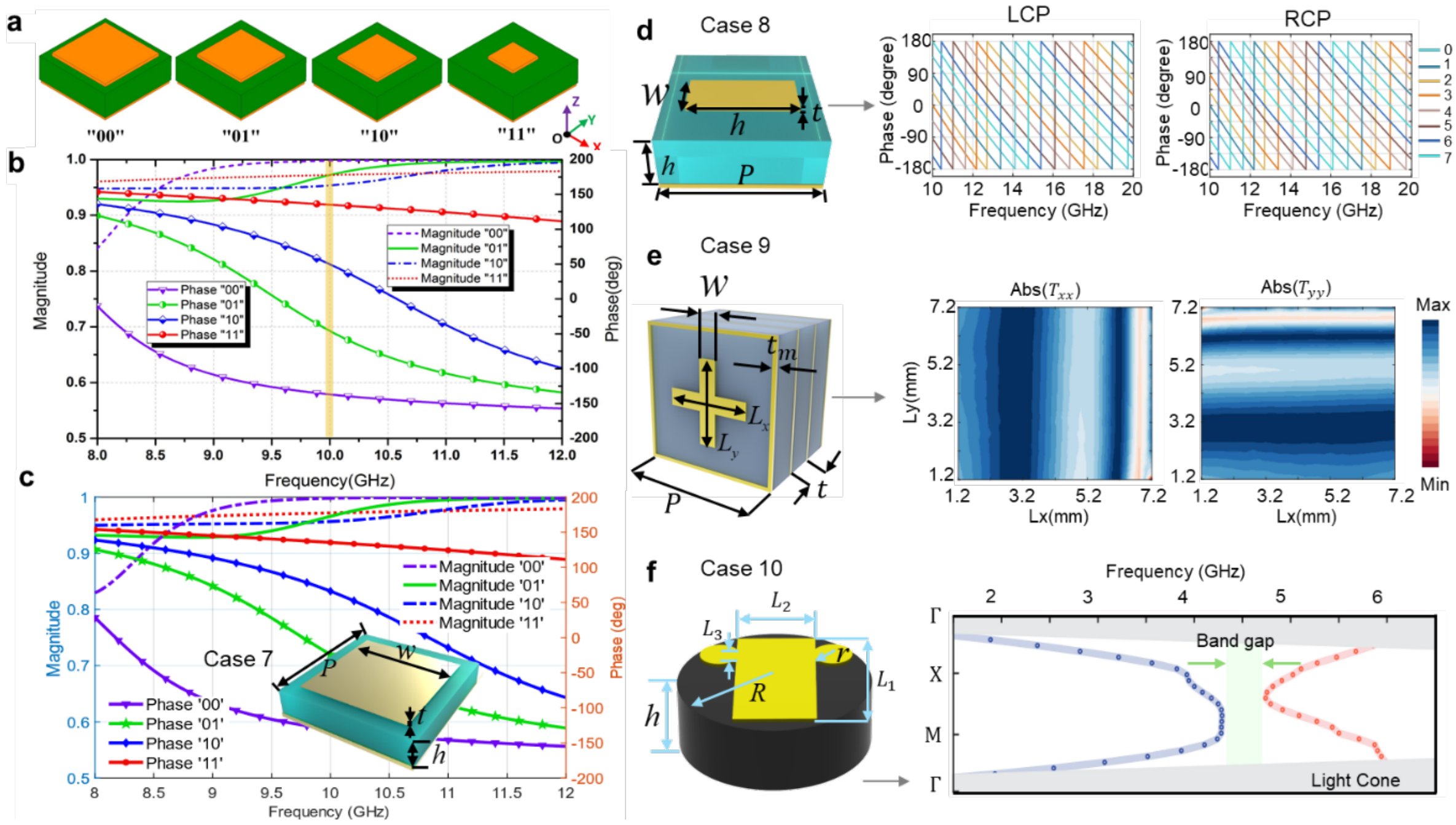

**Fig. 4|** Representative meta-atoms generated by MetaDataGenAgent and their corresponding structure-response results. **a**, Original structural configurations of the polarization-insensitive meta-atom. **b**, Reference simulated magnitude and phase responses from the original literature. **c**, Generated magnitude and phase responses under identical simulation conditions, with the inset showing the corresponding three-dimensional structure. **d**, Generated structural schematic of the PB phase meta-atom and its corresponding phase responses under LCP and RCP incidences. **e**, Generated structural schematic of the anisotropic polarization-sensitive meta-atom and the two-dimensional sweep maps of the co-polarized transmission magnitudes $T_{xx}$ and $T_{yy}$ under varying arm lengths $L_x$ and $L_y$. **f**, Generated structure of the topological photonic unit cell and the corresponding dispersion band structure.

A polarization-insensitive programmable meta-atom reported in ref. 42 is shown in **Fig. 4a** (Case 7), and its reference simulated magnitude and phase responses are presented in **Fig. 4b**. Under identical simulation conditions, the magnitude and phase responses generated by MetaDataGenAgent are plotted in **Fig. 4c**. The generated responses agree closely with the benchmark results reported in the original literature, indicating that MetaDataGenAgent can recover the key geometric parameters required for discrete-state programmable meta-atoms. Specifically, the extracted linewidths are $w = 9.6, 8.1, 7.3$, and 4.6 mm for the "00", "01", "10", and "11" states, respectively, with 90° phase differences between adjacent states. The framework was further evaluated using a PB phase meta-atom reported in ref. 43(Case 8), which represents a different phase-control mechanism based on geometric rotation under circularly polarized incidence. As shown in **Fig. 4d**, the reproduced phase responses recover the expected linear phase progression of the 3-bit meta-atom over the 10-20 GHz band under both left-handed and right-handed circularly polarized incidences. An approximately full 360° phase range is obtained, confirming that MetaDataGenAgent can capture the geometric-rotation and phase-encoding rules associated with PB metasurfaces. Detailed comparisons are provided in Supplementary Note 6.

An anisotropic polarization-sensitive meta-atom reported in ref. 44 (Case 9), corresponding to the cross-shaped resonator shown in **Fig. 4e**, was used to assess performance in a multivariable EM design space. Through multimodal parsing, MetaDataGenAgent correctly identifies the two orthogonal arm lengths, $L_x$ and $L_y$, as independent control variables. The two-dimensional sweep maps in **Fig. 4e** present the generated co-polarized transmission magnitudes, $T_{xx}$ and $T_{yy}$, as the functions of such two geometric parameters. The resulting response distributions agree well with those reported in the original literature and reproduce the decoupled modulation behavior of the $x$- and $y$-polarized waves in the relevant frequency range.

Finally, MetaDataGenAgent was tested on a topological photonic unit cell reported in ref. 14 (Case 10), which involves a more complex modeling task with a dispersion band structure as the target response. As shown in **Fig. 4f**, accurate reconstruction requires not only the dimensions of individual metallic elements, but also their relative placement within the unit cell. Because the original description does not explicitly specify the positions of the two cylindrical patches with respect to the rectangular patch, MetaDataGenAgent identifies this missing geometric information and uses visual structural analysis to estimate their relative positions from the reported schematic. These positions are parameterized in the FDTD model, allowing subsequent adjustment and refinement. The reproduced dispersion diagram resolves the bandgap and light-cone boundaries, showing consistency with the expected topological photonic behavior. Although eigenmode simulation and solver configuration still require limited expert supervision in the present implementation, this example demonstrates that MetaDataGenAgent can reconstruct complex topological meta-atoms from incomplete multimodal descriptions while retaining uncertain parameters for verification and refinement. The representative reconstruction workflow is summarized in Supplementary Note 7, and the detailed comparison between the generated unit cell and the original structure is provided in Supplementary Note 8.

**Experimental validation**

Having demonstrated high-fidelity structure-response generation for individual meta-atoms, we next examined whether the generated meta-atoms could be transferred to device-level electromagnetic functions. Four functional metasurfaces were assembled for beam deflection, holographic imaging, beam focusing, and topological surface-wave transmission. The fabricated metasurface sample is shown in **Supplementary Fig. 4**.

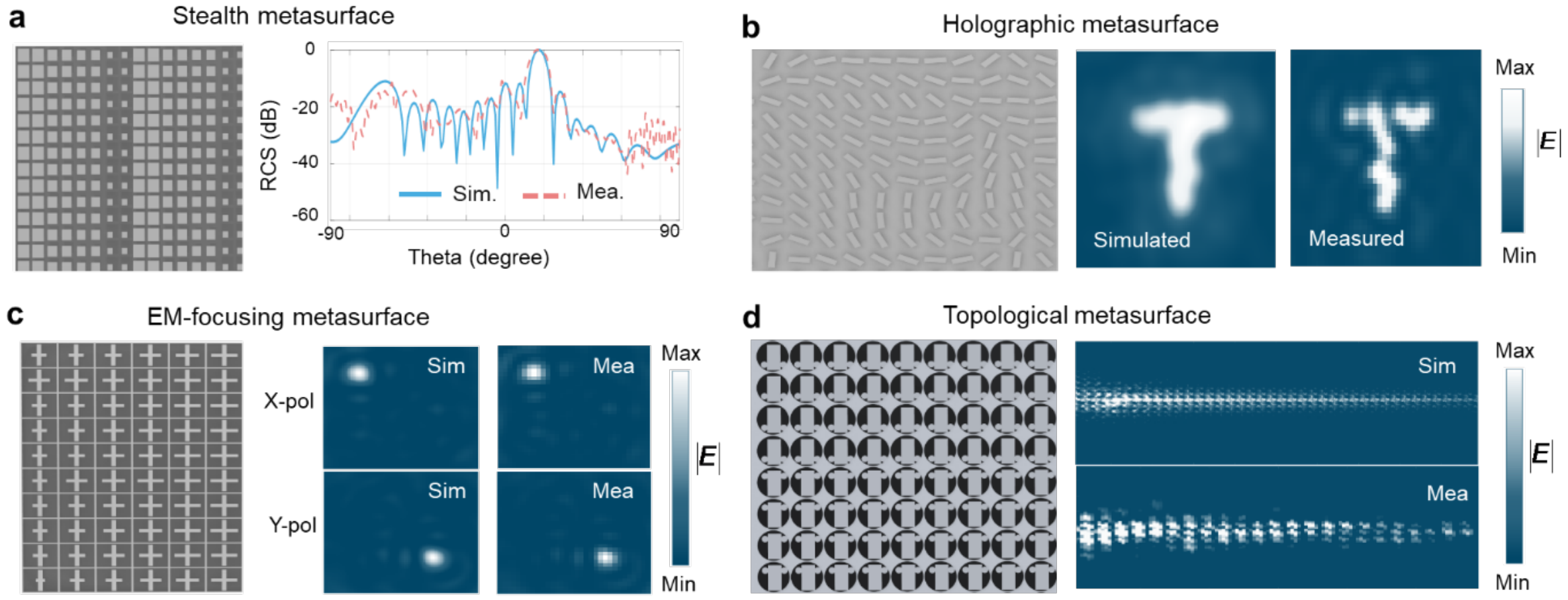

**Fig. 5| Performance of functional metasurfaces composed of meta-atoms generated by MetaDataGenAgent. a**, Coding configuration of the stealth metasurface and its RCS scattering pattern. **b**, Coding pattern of the holographic metasurface and the corresponding simulated and measured holographic images at $z = 500$mm under LCP incidence at 15 GHz. **c**, Coding pattern of the EM-focusing metasurface and the corresponding simulated and measured near-field intensity distributions $|E|$ under the $x$- and $y$-polarized incidences at 12.6 GHz. **d**, Domain-wall configuration of the topological metasurface and the corresponding simulated and measured near-field electric-field distributions $|E_z|$ demonstrating topological surface-wave transmissions at 4.35 GHz. Sim and Mea denote simulations and measurements, respectively.

A stealth metasurface encoded with the "00112233" sequence is presented in **Fig. 5a** together with its far-field radar cross-section (RCS) scattering pattern. The far-field scattering function is given in **Supplementary equation S1**, and the theoretical elevation angle of the reflected beam, calculated from **Supplementary equation S2**, is approximately 17.66°. Both simulation and experiment show that the incident wave is redirected to the designed angle, with good agreement between the measured and simulated scattering patterns. The holographic metasurface is shown in **Fig. 5b**. A modified Gerchberg-Saxton (GS) algorithm[14] was used to determine the phase distribution of the constituent meta-atoms and to generate the coding pattern corresponding to the target holographic pattern "T". Under the illumination by an LCP incident wave at 15 GHz, the holographic image is reconstructed at a plane of $z = 500$mm. The measured holographic image agrees well with the simulated result, although the metasurface consists of only $20 \times 20$ pixelated meta-atoms. These results further support the reliability of the generated meta-atoms in holographic metasurface applications.

Device-level validation was further carried out using an EM-focusing metasurface, as presented in **Fig. 5c**. The coding pattern was designed to realize dual-focus functionality, where the $x$-polarized incident wave is focused at $(-60, 60, 300)$mm and the $y$-polarized incident wave is focused at $(60, -60, 300)$mm. Under illumination at 12.6 GHz, the simulated and measured near-field intensity distributions $|E|$ show good agreement for both polarizations, confirming accurate polarization-dependent focusing at the designed spatial positions. The performance of MetaDataGenAgent in a more complex topological scenario is illustrated in **Fig. 5d**. The generated chiral meta-atoms with opposite topological valley Chern numbers were assembled into a domain-wall metasurface, and the corresponding near-field electric-field distributions $|E_z|$ at 4.35 GHz were obtained in both simulation and experiment. In both cases, the surface waves remain tightly confined to the predesigned domain-wall channel, indicating robust guided transport along the topological interface. Taken together, these device-level results show that the meta-atoms generated by MetaDataGenAgent can support not only accurate meta-atom-level response generation but also the realization of functional metasurface devices with good agreement between simulation and experiment. Details of the near-field and far-field experimental measurements are provided in Supplementary Note 9.

## Discussion

In this paper, we propose MetaDataGenAgent, an orchestrator-driven autonomous multimodal agentic framework for constructing executable electromagnetic metamaterial databases from scientific literature. By integrating multimodal parameter extraction, physics-guided validation and topology analysis, and solver-executable encoding, MetaDataGenAgent transforms unstructured multimodal literature inputs into validated structure-response entries and simulation-ready models. Across representative meta-atoms—including discrete-state programmable, PB phase, anisotropic polarization-sensitive, and topological photonic designs—the framework generates high-fidelity responses that agree with published or benchmarked electromagnetic behaviors. Device-level assembly of the generated meta-atoms further validates their use in functional metasurfaces for beam deflection, holographic imaging, electromagnetic focusing, and topologically protected surface-wave transport. A key advantage of MetaDataGenAgent is its ability to handle incomplete or partially specified structural descriptions. When geometric information is missing from the text, the framework uses visual structural analysis to infer topology and relative element placement, while preserving these quantities as adjustable parameters in the executable model. This capability enables literature-derived structures to remain both simulation-ready and refinement-compatible, which is essential for reliable database construction from heterogeneous scientific reports.

The present implementation remains at a proof-of-concept stage. Highly complex non-standard three-dimensional structures, irregular cross-scale topologies, and sketches with insufficient geometric views may still challenge current multimodal models in spatial interpretation and fine-feature recovery. Nevertheless, the modular and model-extensible architecture of MetaDataGenAgent allows future advances in multimodal reasoning, visual grounding, and code generation to be directly incorporated into the framework. Looking forward, MetaDataGenAgent provides a practical basis for integration with inverse design algorithms and for extension to broader classes of metamaterials, including acoustic, thermal, and multiphysics systems. More broadly, this work suggests a general route for converting knowledge-rich scientific literature into executable, AI-ready databases for data-scarce research fields. This work therefore has implications not only for metamaterial research but also for the emerging field of AI for science.

## Methods

### Structure design and measurements

The metasurfaces designed in this study were fabricated using printed circuit board (PCB) technology. The dielectric substrates for the meta-atoms shown in Figures 4c-e are F4B, whereas that for the meta-atom shown in Figure 4f is FR4. Figure 5 shows the structural or coding configurations of the corresponding metasurfaces. For full-wave simulation and experimental validation, the associated metasurface arrays were implemented with a size of $20 \times 20 = 400$ meta-atoms. Figure 5d presents the configuration of the topological metasurface, while the corresponding array used in simulation and experiment consisted of $16 \times 20 = 320$meta-atoms. Details of the near-field and far-field experimental measurements are provided in Supplementary Note 9.

### Numerical simulations

All numerical simulations in this study were performed using the commercial software CST Microwave Studio and COMSOL Multiphysics. The simulated electric field distributions for wave manipulation were obtained using the time-domain solver in CST Microwave Studio. These simulation results further corroborate the accuracy of the unit cells reproduced by MetaDataGenAgent, demonstrating the universality of this approach and the feasibility of its practical applications.

### Foundation model configurations

MetaDataGenAgent was implemented using a multimodal foundation model that integrates both textual and visual reasoning capabilities, enabling the joint interpretation of literature text, captions, and structural schematics. The model is used to extract geometric, material, and simulation-related information, perform physical-logic validation, and generate solver-executable modeling instructions for electromagnetic simulations. By processing textual and visual inputs within a single framework, the model allows MetaDataGenAgent to recover complex structural topology and dimensional information even when certain geometric parameters are not explicitly specified. Streaming output was enabled to support long-response generation. Detailed agent architectures, model settings, and prompt-engineering strategies are provided in Supplementary Notes 1-5.

## Acknowledgements

The work was supported by the National Key Research and Development Program of China (2023YFB3813100), National Natural Science Foundation of China(62288101, 61925103), Special Fund for Key Basic Research in Jiangsu Province (BK20243015), State Key Laboratory of Millimeter Waves, South-east University, China (K201924), 111 Project (111-2-05).

### Author Contributions

S.Q. developed the MetaDataGenAgent framework, performed the main computational experiments, organized and analysed the data, and wrote the manuscript. Z.Y. designed the metasurface experiments, fabricated the samples, performed the physical measurements, and collected and organized the experimental data. X.Z. contributed to computational experiments and data organization. Y.Z. assisted with computational experiments. A.Y. contributed to data collection. K.Z. and Q.C.C. contributed to literature research and background analysis. J.W.Y. and T.J.C. developed the MetaDataGenAgent framework, conceived and supervised the project, guided the experimental design and data analysis, and revised the manuscript. All authors discussed the results and commented on the manuscript.

### Competing interests

The authors declare no competing financial interests.

### Data and materials availability

The data that underpin the findings of this study are available from the corresponding author on reasonable request. All code generated and used for the analyses in this study can be retrieved from the GitHub repository: https://github.com/qinshilong0119/MetaDataGenAgent.

**Supporting Information**

Supporting Information is available from the Online Library or from the author.